\documentclass[seceq]{ptptex}

\usepackage{graphicx}
\usepackage{wrapft}

\title{
Phenomenological Analogies in Black Hole Systems of all Masses
}

\author{
I.F. \textsc{Mirabel}$^{1,2}$%
}

\inst{
$^1$Service d'astrophysique. CEA-Saclay. France\\
$^2$Instituto de Astronom\'ia y F\'isica del Espacio. CONICET. Argentina\\
}

\abst{
I review the progress made on the physics of relativistic jets from black 
hole systems in the context of the analogy between 
AGN and microquasars that was proposed one decade ago. 
If the emerging empirical correlations between the observational properties 
of stellar and supermassive black holes will become more robust, 
we will use them to determine the mass and spin of black holes, 
independently of theoretical models.  Microquasars are fossils of 
sources of Gamma-ray bursts (GRBs) of long duration, and their kinematics 
provides observational clues on the physics of collapsars.  
If jets in GRBs, microquasars and AGN are due to a unique universal 
magnetohydrodynamic mechanism, synergy between the research on these three 
different classes of cosmic objects will lead to further progress in 
black hole physics and astrophysics. 

}

\begin{document}

\maketitle

\section{Physics in black holes of all mass scales}

The physics in all systems dominated by black holes is essentially the same, 
and  it is governed by the same scaling laws. 
The main differences derive 
from the fact that the scales of length and time of the phenomena 
are proportional to the mass of the black hole. If the lengths, 
masses, accretion rates, and luminosities are expressed in 
units such as the gravitational radius 
(R$_g$ = 2GM/c$^2$), the solar mass, 
and the Eddington luminosity, the same physical laws apply to 
stellar-mass and supermassive black holes\cite{Sams,Rees}. 
For a black hole of mass M the density and mean temperature in the accretion flow scale 
with M$^{-1}$ and  M$^{-1/4}$, respectively. For a given
critical accretion rate, the bolometric luminosity and length of relativistic jets are proportional to the mass of the black hole. 
The maximum magnetic field at a given radius in a radiation dominated accretion disk scales with  M$^{-1/2}$, which implies that 
in the vicinity of stellar-mass black holes the magnetic fields may 
be 10$^4$ times stronger 
than in the vecinity of supermassive black holes\cite{Sams}. 
In this context, it was proposed\cite{Mirabelnat92,Mirabelnat98} that supermassive black holes in quasars and stellar-mass black holes in 
x-ray binaries should exhibit analogous phenomena. Based on this 
physical analogy, the word 
``microquasar"\cite{Mirabelnat92} was chosen to designate compact 
x-ray binaries that are sources of relativistic jets (see Figure 1). 

\begin{figure}
\centerline{
\includegraphics[width=14cm]{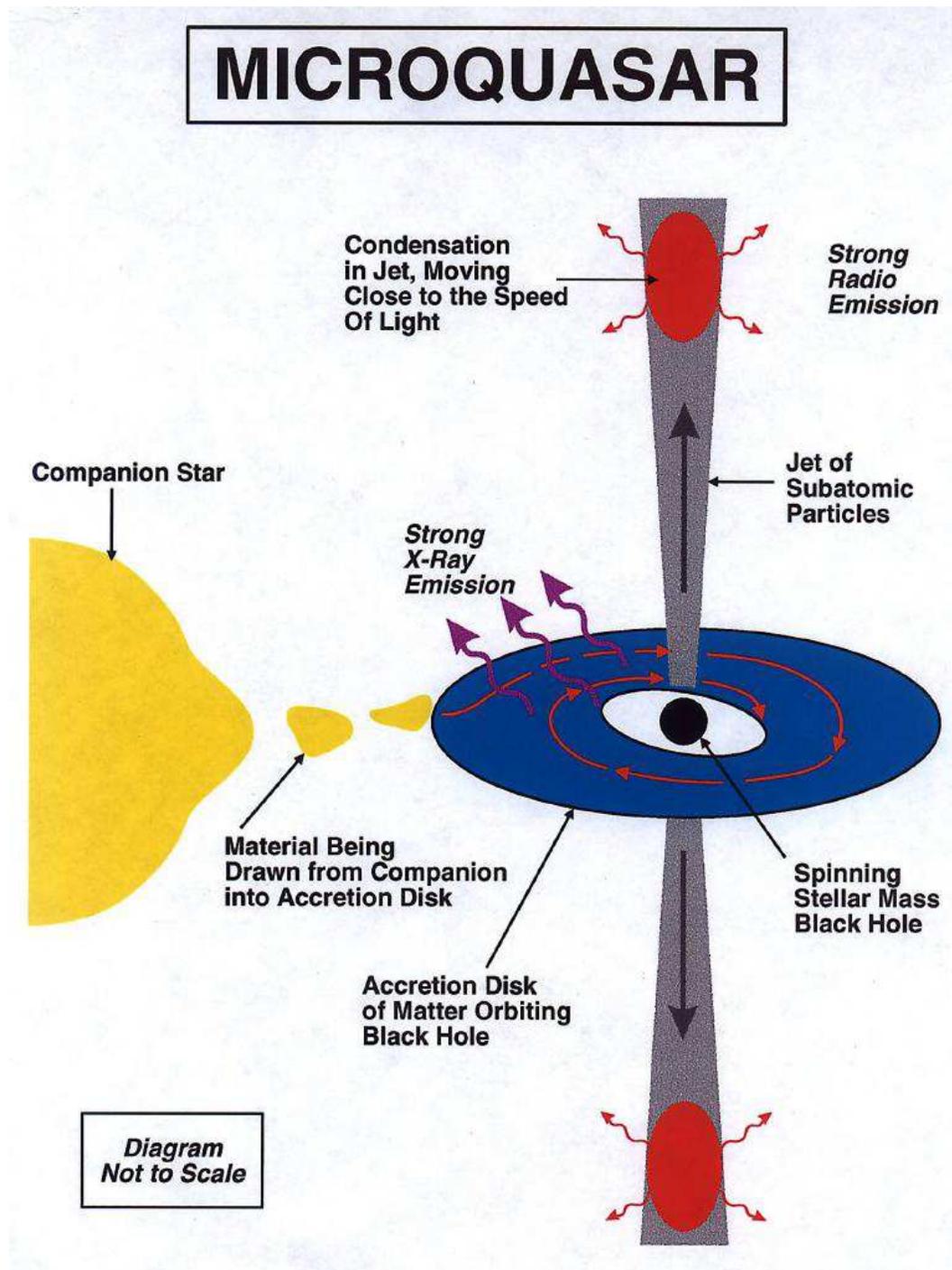}}
\caption{This diagram illustrates current ideas of what may be microquasars.  
These are x-ray binary systems that eject plasma at relativistic speeds.}
\label{mylabel1}
\end{figure}



\section{Superluminal motions in AGN and microquasars}

A galactic superluminal ejection was observed for first 
time in the black hole x-ray binary GRS 1915+105,  
at the time of a sudden drop in the BATSE 20-100 keV flux\cite{Mirabelnat94}. 
Since then, relativistic jets with comparable bulk Lorentz factors 
$\Gamma$ = 1/[(1-$\beta$)$^2$)]$^{1/2}$ as 
in quasars have been observed in several other 
x-ray binaries\cite{MirabelARAA,Fender1,Paredes}. At present, it is believed  
that all x-ray accreting black hole binaries are jet sources. 

\begin{figure}
\centerline{
\includegraphics[width=16cm]{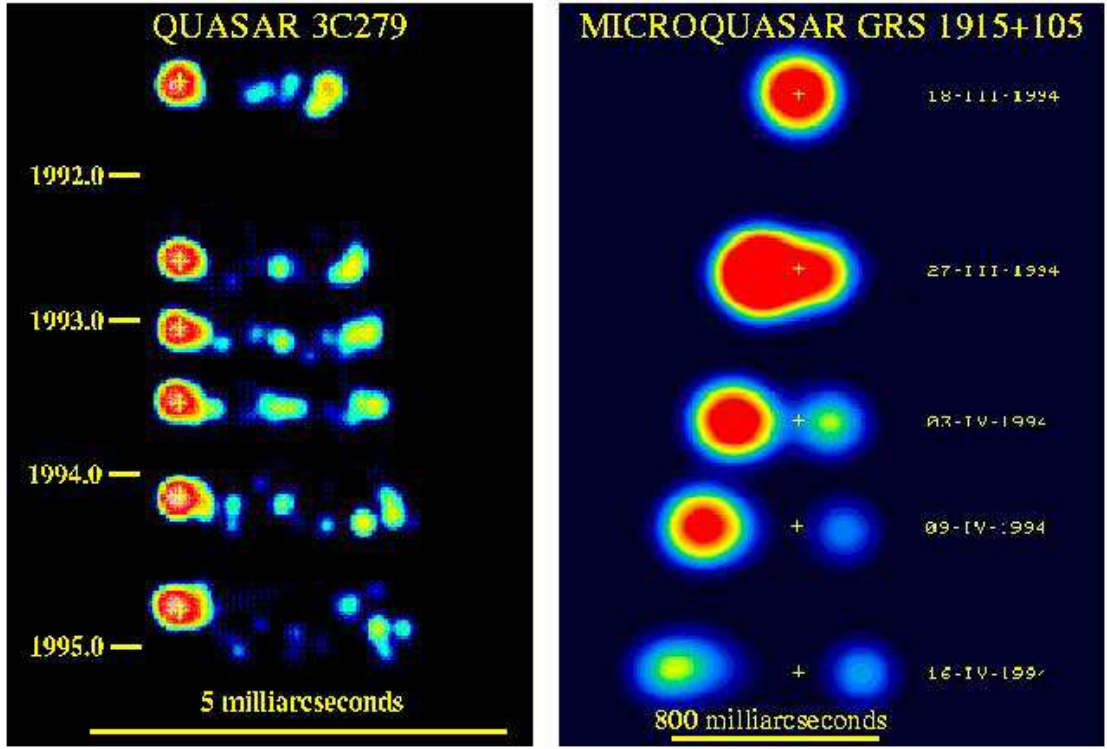}}
\caption{Apparent superluminal motions observed  
in the microquasar GRS 1915+105 at 8.6 GHz\cite{Mirabelnat94} and in 
the quasar 3C 279 at 22 GHz.}
\label{mylabel2}
\end{figure}

Galactic microquasar jets usually move in the plane of the sky 
$\sim$10$^3$ 
times faster than quasar jets and can be followed 
more easily than the later  (see Figure 2). Because of their proximity, in microquasars two-sided jets 
can be observed, which together with the distance provides the necessary data to solve the system of 
equations, gaining insight on the actual speed of the ejecta. 
On the other hand, 
in AGN located at 
$\leq$100 Mpc, the jets can be imaged with resolutions of a few times 
the gravitational radius of the supermassive black hole, as was done for 
M 87\cite{Biretta}. This is not 
presently possible in microquasars, since such a precision in terms of 
the gravitational radius of a stellar-mass black hole would 
require resolutions a few hundreds of kilometers. Then, in terms of the 
gravitational radius
in AGN we may learn better  
how the jets are collimated close to the central engine. In summary, 
some aspects of the relativistic jet phenomena associated to accreting 
black holes are better observed in 
AGN, whereas others can be better studied in microquasars. 
Therefore, to gain 
insight into the physics of relativistic jets in the universe, synergy 
between knowledge of galactic and extragalactic black hole is needed.

\section{Accretion-jet connection in microquasars and quasars}

Microquasars have allowed to gain insight into the connection 
between accretion disk instabilities and the formation of jets.  
In $\sim$1 hour of simultaneous multiwavelength observations of GRS 1915+105 during the frequently observed 30-40 min x-ray oscillations in this 
source, the connection between sudden drops of 
the x-ray flux from the accretion disk and the onset 
of jets were observed in several ocassions\cite{MirabelAA98,Eikenberry} 
(see Figure 3).

\begin{figure}
\centerline{
\includegraphics[width=14cm]{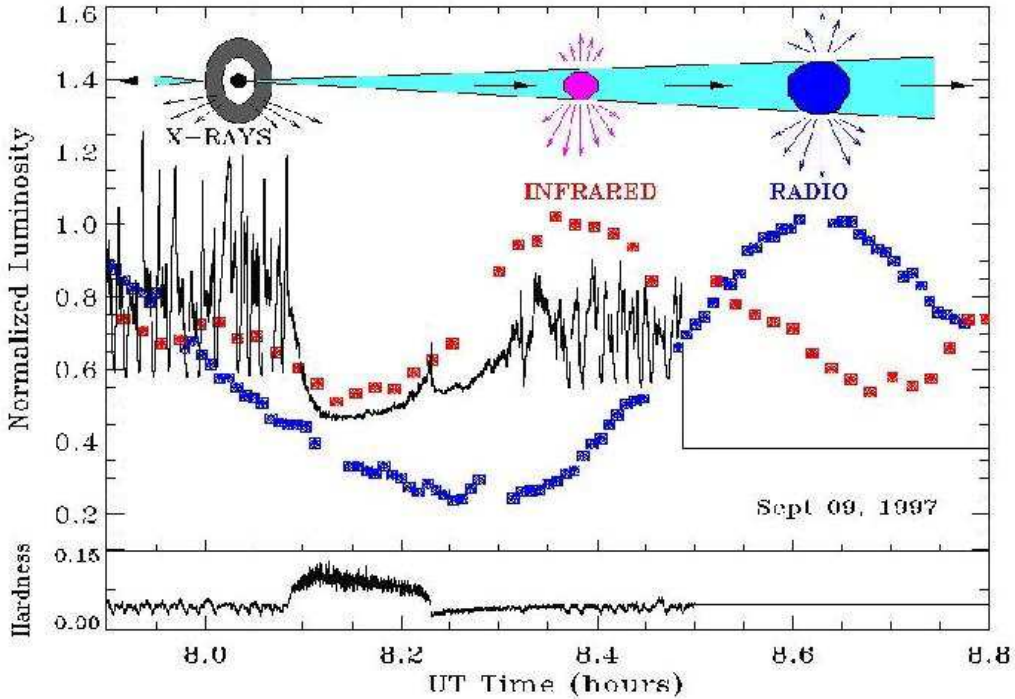}}
\caption{Direct evidence for the disk-jet connection in the black hole x-ray 
binary GRS 1915+105\cite{MirabelAA98}. When the hot inner accretion disk disappeared, its x-ray brightness abruptly diminished. The ensuing x-ray recovery documented the inner disk's replenishment, while the rising infrared and radio emission showed plasma being ejected in a jet-forming episode. 
The sequence of events shows 
that material indeed was transfered 
from the disk to the jets. Similar transitions have been observed in the 
quasar 3C 120\cite{Marscher}, but in time scales of years, rather than minutes.}
\label{mylabel3}
\end{figure}

From these observations we have learned the following:

a) the jets appear after the drop of the x-ray flux, 

b) the jets are produced during the replenishment of the inner 
accretion disk, 
 
c) the jet injection is not instantaneous. It can last up to 
$\sim$10 min,
 
d) the time delay between the jet flares at wavelengths of 2$\mu$m, 
2cm, 3.6cm, 6cm, and 21cm are consistent with the model of 
adiabatically expanding clouds that had been proposed to account 
for relativistic jets in AGN\cite{vanderLaan},

e) synchrotron emission is observed up to infrared wavelengths and 
probably up to x-rays. This would imply the presence in the jets of electrons with TeV energies,

f) VLBA images during this type of x-ray oscillations\cite{Dhawan} showed that 
the ejecta consist on compact collimated jets with lengths of $\sim$100 AU.

g) there is a time delay of $\sim$5 
min between the large drop of the x-ray flux from the accretion 
disk and the onset of the jets. These $\sim$5 minutes of silence suggest 
that the compact object in GRS 1915+105 has a space-time border, rather than a material border, namely, a horizon as expected in 
relativistic black holes. However, the absence of evidence of 
a material surface in these observations could have alternative explanations.

After the observation of this accretion disk-jet connection in a microquasar, 
an analogous connection was 
observed in the quasar 3C 120\cite{Marscher}, but in scales of 
years rather than minutes. This time scale ratio is 
comparable to the mass ratio between the supermassive black 
hole in 3C 120 and the stellar black hole in GRS 1915+105, as expected 
in the context of the black hole analogy. 

\section{X-ray jets in AGN and microquasars}

X-ray emission has been observed with Chandra and XMM-Newton in the 
radio jets, lobes and hot spots of quasars and radio galaxies. 
X-ray photons 
can be produced by inverse Compton scattering from the environment of 
the central engine up to distances of $\sim$100 kpc\cite{Wilson}. 
Synchrotron x-ray radiation has been 
detected in some sources, most notably in the jet of M 87. 

Steady, large-scale radio jets are associated to  
x-ray persistent microquasars\cite{MirabelARAA}.  
Extended x-ray emission associated to the radio emission was observed in 
the galactic source SS433/W50 up to distances of $\sim$30 pc from 
the central engine\cite{Dubner} (see Figure 4). The x-rays extend  
in the same direction as the sub-arcsec, precessing jets. 
The jets in SS433 are hadronic, move with a velocity of 0.26c, 
and have a kinetic power of $\sim$10$^{39-40}$ erg s$^{-1}$. 
Since no hot spots have 
been detected in the shock regions 30 pc away from the central 
engine, SS433/W50 cannot be considered a scale-down analog of a 
FR II radio galaxy. Most microquasars with extended jet emission 
have morphologies analogous to FR I's rather than FR II's radiogalaxies. 

\begin{figure}
\centerline{
\includegraphics[width=14cm]{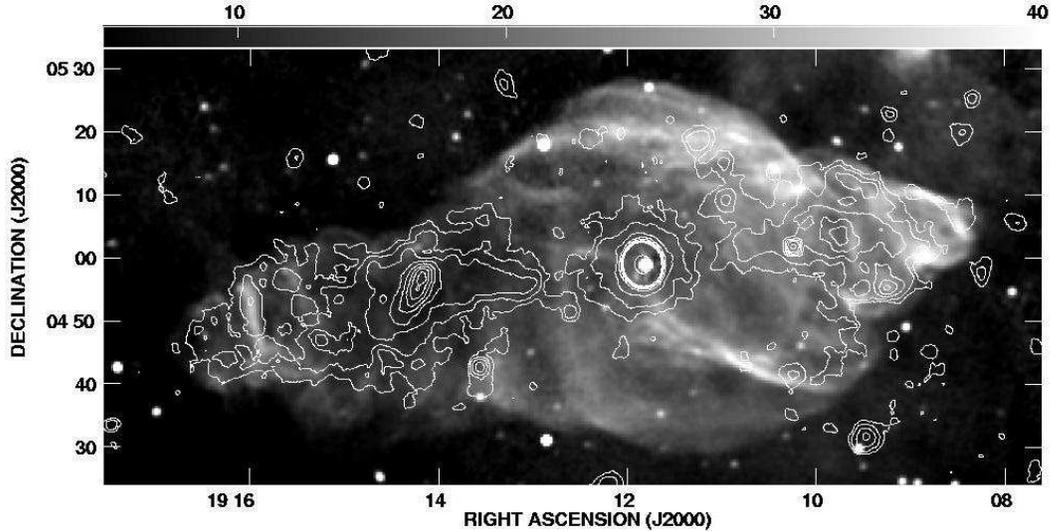}}
\caption{VLA radio image at 1.5 GHz (grey scale) superimposed on the ROSAT 
x-ray contours of SS433/W50\cite{Dubner}. The radio counterpart of the microquasar is 
the bright unresolved source at the center of the image. The lateral E-W extension of the nebula over $\sim$1$^{\circ}$ ($\sim$50 pc) is caused by 
the injection of relativistic jets from SS 433.}
\label{mylabel4}
\end{figure}

The recent discovery of radio\cite{Hjellming} and x-ray\cite{Corbel} 
moving jets in  
microquasars rise the possibility of studying the formation 
of radio and x-ray lobes in real time. These observations show that jets 
may transport energy in a ``dark" way, namely, in a way that is radiatively  inefficient, until shocks are produced. Synchrotron x-ray emission 
from shocks at large distances from the central 
engines imply that microquasars are potential sources of cosmic rays and 
electrons with up to TeV energies.

\section{Blazars and microblazars}

The bulk Lorentz factors $\Gamma$ in microquasars and quasars 
have similar values, 
and it had been proposed\cite{MirabelARAA} that microquasars with 
jet axis that form angles $\leq$10$^{\circ}$ with the line of sight 
should appear as ``microblazars", showing analogous phenomena
to blazars. Due to relativistic beaming, 
in microblazars the brightness of outbursts is enhanced by factors of 
8 $\times$ $\Gamma$$^3$ and the interval of time of the phenomena 
is reduced 
by factors of 1/2 $\times$ $\Gamma$$^{-2}$. Then, microblazars 
should appear as intense sources 
of high energy photons with very fast variations of flux, which makes 
them difficult to find and to follow. Due to this difficulty and to 
the relatively low statistical probability of small angles between the jet axis 
and the line of sight\cite{MirabelARAA}, it is not surprising that most 
of the 
microquasars studied so far exhibit large angles ($\geq$30$^{\circ}$) between the jet axis and the line of sight.

It has been proposed that microblazars may be more frequently  
found in High Mass X-ray Binaries (HMXBs)\cite{Paredes2,Romero}. 
In such binaries, gamma-rays can be produced by inverse Compton 
of the jet particles with the UV photons radiated by the massive donnor 
star. In fact, the three microquasars so far proposed as counterparts 
of variable EGRET unidentified sources are HMXBs with similar 
properties\cite{Combi}.

\section{Extragalactic microquasars and super-Eddington x-ray sources}

GRS 1915+105 and SS 433 may be the Milky Way counterparts 
of the two  classes of most numerous super-Eddington x-ray sources 
found in external galaxies\cite{King}. GRS 1915+105 is a long lasting 
transient outburst x-ray binary with an evolved donnor of $\sim$1 M$_{\odot}$, 
whereas SS 433 is a persistent HMXB. SS 433 type of ULX's are preponderantly 
found in starburst galaxies like the Antennae, whereas luminous x-ray 
sources of low mass as GRS 1915+105 may also be found in galaxies with 
a low rate of star formation. 

Most of the ULX's would be stellar-mass black hole microquasars with the following possible properties:
 
1) HMXBs that host massive stellar black holes (M $\geq$ 40 M$_{\odot}$) with  isotropic radiation\cite{Pakull}.

2) HMXBs and LMXBs that host stellar black holes (M $\sim$ 10 M$_{\odot}$) 
with anisotropic radiation\cite{King}. 

3) A few ($\leq$1\%) may be microquasars with relativistic boosted radiation. 
These should be very bright, highly time-variable, and have a hard x-ray/$\gamma$-ray photon spectrum.\cite{MirabelARAA}.

Although less numerous, it is not exculded that some ULX's could be 
accreting black holes of intermediate-mass (100-1000 M$_{\odot}$).

\section{X-ray/Radio correlations in low power black holes of all masses}

Several teams of researchers are exploring interesting 
x-ray/radio correlations. 

Microquasars in the low-hard state exhibit radio/x-ray 
correlations\cite{Gallo}. In the low-hard state the power output 
of quiescent black holes  
is jet-dominated and when the system moves to a high soft state 
the radio jets are quenched. The same seems to take place 
in AGN\cite{Maccarone}.     
A scheme to unify low-power accreting black holes has 
been proposed\cite{Falcke}, where the black holes in Sgr A$^*$, 
LINERs, FR I, and BL Lac would be analogous to microquasar black holes 
in the low-hard state. 

Following studies\cite{Heinz} of correlations between radio and 
bolometric luminosities, a fundamental plane of black hole activity in terms of the black hole 
mass and x-ray and radio core luminosities is 
proposed\cite{Merloni}. This correlation holds 
for radiativelly inefficient accretion, not for bright thin 
synchrotron emitting states. 

At present, these empirical correlations have large scatters. 
However, if they became more robust, the mass of black holes 
could be inferred from the x-ray and radio fluxes, 
independently of theoretical models.

\section{Time variations of flux and the masses of black holes}

Time variations of flux may be correlated  
with the mass of the black hole. 

1) The duration of the x-ray flares observed in stellar-mass black 
holes and in Sgr A* seem to be proportional to the mass of the black 
holes. In Cygnus X-1 and other x-ray black hole binaries, flares with 
durations of 1-10 ms are observed\cite{Gierlinski}. On the other 
hand in Sgr A$^*$, x-ray flares lasting 400-10,000 sec have been 
observed with Chandra\cite{Baganoff} and XMM-Newton\cite{Goldwurm}. As 
expected, the time ratios of the power variabilities are comparable to the black hole mass ratios. 

2) For a given black hole spin, the  maximum frequencies of quasi 
periodic oscillations (QPOs) of flux are expected to be proportional 
to the mass of the black hole. In 4 microquasars, 3:2 twin 
peak x-ray QPOs 
of maximum frequency in the range of 100-500 Hz have been observed, from which angular momenta a = J/(GM/c$^2$) = 0.6-0.9 
have been derived\cite{Abramowicz}.  On the other hand, 17 min infrared 
QPOs have been reported in Sgr A$^*$, from which it has 
been inferred an angular momentum a = 0.52\cite{Genzel}.  
As expected, these QPOs appear to scale with the mass of the black 
hole. If the 17 min QPO in Sgr A$^*$ is confirmed as a 
component of a twin peak fix QPO of maximum frequency, 
this correlation could be used to 
derive black hole masses, and in particular, those of the super-Eddington 
x-ray sources in external galaxies\cite{Abramowicz}. 

3) Some properties of the aperiodic variability (noise) in AGN and x-ray 
binaries seem to be correlated with the mass of the compact objects. 
The break time scale in the power spectra density of black holes 
seems to scale linearly with the mass of the black hole\cite{Uttley}. 
The broad band break time in the Sey 1 NGC 3516 scales linearly with that 
of Cyg X-1 in the 
low-hard state\cite{McHardy}. If this type of correlation is confirmed it 
could also be used to estimate the mass of black holes 
in extragalactic super-Eddington x-ray sources.

\section{Relativistic iron lines in stellar and supermassive black holes}

AGN frequently exhibit broad iron K$\alpha$ lines skewed to low 
energies\cite{Tanaka}. The shape of these lines is consistent 
with emission from the surface of an accretion disk extending from about 
6 to more than 40 gravitational radii. Occasionaly the red wing of the 
line extends below 4 keV and the current explanation is that the disk 
extends within 6 gravitational radii implying that the black hole is rapidly spinning. Now it is widely believed that this spectral feature 
is a probe of the immediate environment of black holes\cite{Fabian}. 
Until recently, only smeared edges with little evidence for line emission 
had been observed in Galactic black hole binaries. But after 
Chandra, XMM-Newton and Beppo-SAX, similar emission iron lines 
to those in AGN were found, even in the ASCA archive\cite{Miller}. 

Besides emission lines skewed to low energies, analogous spectra to AGN-like warm 
absorbers  are observed in some x-ray binaries\cite{Miller2}. 
The absorption is variable and it is believed to be produced in a 
dense local disk wind rather than in the ISM. A finding possibly 
related to these absorption lines is the discovery with INTEGRAL of black 
hole binaries with strong x-ray 
absorption, much larger than that derived from  
optical and infrared observations, which also implies that the absorption 
is local rather than in the ISM. 

The iron line in stellar black hole binaries can be used 
to investigate: 

1) the physical models of the K$_{\alpha}$ line. Because of the short 
dynamical time scales, the shape of the line can be correlated 
with the x-ray state of the accretion disk, and corona-disk interactions.
 
2) the dense plasma outflows from accreting black holes. 
The study of warm absorber lines  similar to those seen in Seyferts 
may be important to estimate the mass outflows in x-ray binaries. 

3) the spin of the accreting black hole. This is important to test 
models where the jets are powered by the spin of the black hole.

At present, the main constrain to derive the slope of the iron lines is due to the uncertainties on the shape of the continuum at energies 
$\geq$8 keV.

\section{Microquasars as fossils of gamma-ray burst sources}

Gamma-ray bursts (GRBs) of long duration are believed to be jets from 
collapsars at cosmic distances. In this context there should 
be astrophysical and physical connections between GRB sources 
and microquasars. If GRB sources are microquasars 
in formation at cosmic distances, microquasars in our galaxy 
should contain clues on the physics in collapsars. 

It is believed that collapsars that produce GRBs take place in close massive binaries because: 

1) the core must be spun up by spin-orbit interaction in order to provide 
enough power to the jet that drills  the collapsing star 
all the way from the core up to the external layers\cite{Izzard}.  

2) GRBs seem to be associated to SNe Ic. This is the class of SNe that 
do not show H and He lines, implying that before the explosion 
the progenitor of those GRBs had lost the H and He layers. These 
layers are more easily lost if the progenitor was part 
of a massive binary that underwent a common envelope phase. 
Furthermore, SNe Ic exhibit 4-7 \% polarization, which are an indication 
of asymmetric explosions caused by collimated jets. 

As for AGN, it has been proposed\cite{Lamb,Kouveliotou} an unification 
scheme where GRBs, x-ray flashes and SNe Ic are the same phenomenon, 
but viewed from different angles. 
However, radio observations of SNe Ic  
suggest\cite{Soderberg} that the characteristics of supernovae are 
not dictated by the viewing angle but rather by the properties of the 
central engine. The observed variety of cosmic explosions (GRBs, x-ray 
flashes, SNe Ic) would then be explained by the varying fraction of the 
explosion energy that is channeled into relativistic ejecta. 
  
\section{Constrains on collapsar physics from microquasar kinematics}

It is believed that stellar black holes can be formed in two different
ways: Either the massive star collapses directly into a black hole
without a supernova explosion, or an explosion occurs in a
protoneutron star, but the energy is too low to completely unbind the
stellar envelope, and a large fraction of it falls back onto the
short-lived neutron star, leading to the delayed formation of a black
hole\cite{Fryer}. If the collapsar takes place in a binary that  
remains bound, and the core collapse produced an energetic 
supernova, it will impart the center of mass of the system with a 
runaway velocity, no matter the explosion being symmetric or 
asymmetric. Therefore, the kinematics of microquasars can be used 
to constrain theoretical models of the explosion of massive stars that  
form black holes. 

Recently, the runaway velocities of 
several microquasars have been determined\cite{MirabelSci,MirabelAA2,Ribo}. Although the number statistics 
is still rather low, the preliminary results 
are consistent with evolutionary models for binary massive 
stars\cite{Fryer}, where  neutron stars 
and low-mass black holes 
form in energetic supernova explosions, whereas black holes 
with the larger masses form silently.


\begin{figure}
\centerline{
\includegraphics[width=14cm]{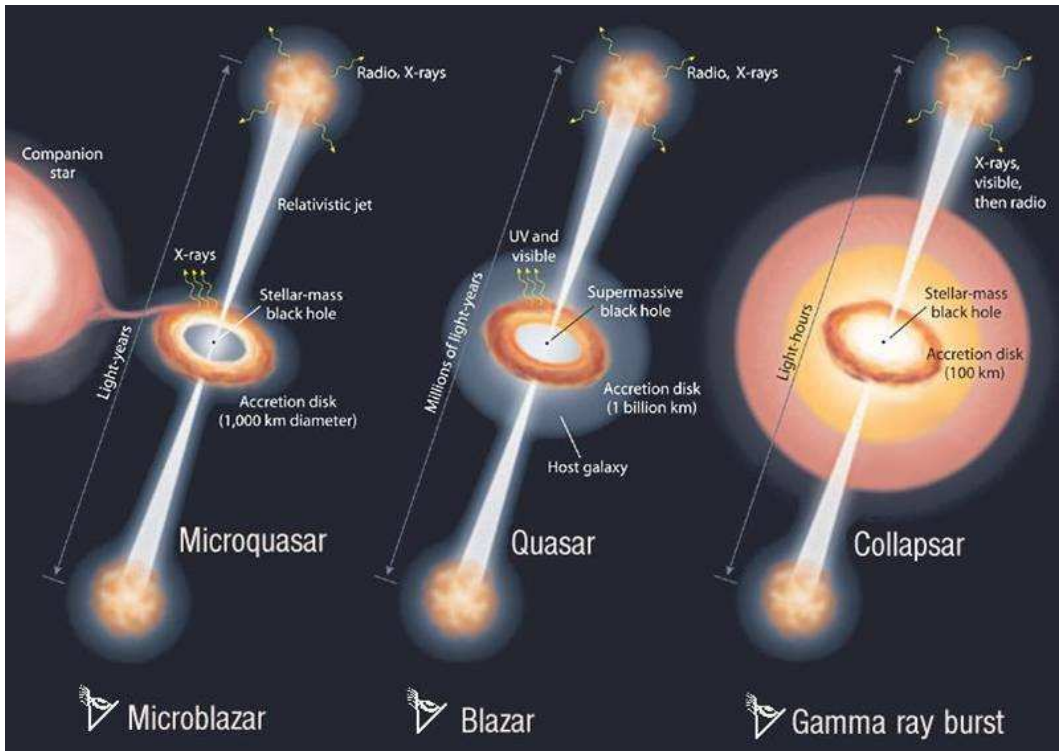}}
\caption{A unique universal mechanism may be responsible for three types of 
astronomical objects: microquasars (left); quasars (center); and 
collapsars (right), the massive, suddenly collapsing stars believed to cause some gamma-ray bursts. Each contains a black hole (probably spinning), an accretion disk (which transfers material to the black holes), and relativistic jets (which emerge from a region just outside the black holes, carrung away angular momentum). Microquasars and quasars can eject matter many times, while collapsars form jets but once. When the jet is aligned with an observer's line of sight these objects appear as microblazars, blazars, and gamma-ray bursters, respectively. The components of each panel are not drawn to scale); scale bars denote jet lengths. (Sky \& Telescope, May 2002, 32)}
\label{mylabel4}
\end{figure}

\section{Relativistic jets in AGN, microquasars and GRB sources}

It was suggested that irrespective of their 
mass there may be a unique universal mechanism for 
relativistic jets in accreting black holes\cite{MirabelST} (see Figure 5). 
Although in AGN, microquasars and GRB sources there are different 
physical conditions, it was proposed  
that all jets are produced by an unique electromagnetic mechanism, in which 
charged plasma is accelerated by electric fields that are generated 
by a rotating magnetic field (Meier, astro-ph/0312047). However, the most popular GRB 
jet models at this time are baryon dominated, 
and the factors that control 
the jet power, collimation and speed, remain unknown.

\section*{Acknowledgements}
I would like to thank Professor K. Makishima for his kind hospitality in Japan. 
I also wish to thank my colleagues  L.F. Rodr\'\i guez, I. Rodrigues, S. Chaty, V. Dhawan,  M. Rib\'o, R. Mignani, D. Wei, J.A. Combi and G.E. Romero for the contribution they have made in the research reviewed here, and L. Pellizza for reading the manuscript.

%

\end{document}